\documentclass[showpacs,aps,prb,twocolumn,superscriptaddress]{revtex4-1}
\usepackage{graphicx} 
\usepackage{dcolumn}
\usepackage{bm}
\usepackage{amssymb,amsmath}
\usepackage{epstopdf}
\usepackage{color}

\begin{document}
\title{Magnetophonon resonance in graphite: High-field Raman measurements and electron-phonon coupling contributions}

\author{Y. Kim}
\affiliation{National High Magnetic Field Laboratory, Tallahassee, FL 32310, USA}
\affiliation{Department of Physics, Florida State University, Tallahassee, FL 32306, USA}
\author{Y. Ma}
\author{A. Imambekov}
\affiliation{Department of Physics and Astronomy, Rice University, Houston, Texas 77005, USA}
\author{N. G. Kalugin}
\affiliation{Department of Materials and Metallurgical Engineering, New Mexico Tech, Socorro, NM 87801, USA}
\author{A. Lombardo}
\author{A. C. Ferrari}
\affiliation{Engineering Department, Cambridge University, Cambridge, CB3 0FA, UK}
\author{J. Kono}
\affiliation{Department of Physics and Astronomy, Rice University, Houston, Texas 77005, USA}
\affiliation{Department of Electrical and Computer Engineering, Rice University, Houston, TX 77005, USA}
\author{D. Smirnov}
\email{smirnov@magnet.fsu.edu}
\affiliation{National High Magnetic Field Laboratory, Tallahassee, FL 32310, USA}

\date{\today}

\begin{abstract}

We perform Raman scattering experiments on natural graphite in magnetic fields up to 45~T, observing a series of peaks due to interband electronic excitations over a much broader magnetic field range than previously reported. We also explore electron-phonon coupling in graphite via magnetophonon resonances. The Raman $G$ peak shifts and splits as a function of magnetic field, due to the magnetically tuned coupling of the $E_{2g}$ optical phonons with the $K$- and $H$-point inter-Landau-level excitations. The analysis of the observed anticrossing behavior allows us to determine the electron-phonon coupling for both $K$- and $H$-point carriers. In the highest field range ($>$35~T) the G peak narrows due to suppression of electron-phonon interaction.

\end{abstract}

\pacs{78.30.-j, 71.70.Di, 73.61.Cw, 76.40.+b, 78.20.Bh}

\maketitle

Electron-phonon coupling in graphene and graphite has been investigated for several years\cite{PiscanecetAl04PRL,FerrarietAl06PRL,Ando06JPSJ2,Ferrari07ssc}. The zone-centre, doubly degenerate, $E_{2g}$ phonon strongly interacts with electrons, resulting in renormalization of phonon frequencies
and line broadening\cite{PisanaetAlNatMat07, Yan07, DasetAl08NatureNano, DasetAl09PRB,YanPRB09, Lazzeri06}. These are tunable by electric and magnetic fields, through Fermi-energy shifts and Landau quantization.  The Raman $G$ peak is predicted to exhibit anticrossings when the $E_{2g}$ phonon energy matches the separation of two Landau levels (LLs).  Both intraband (i.e., cyclotron resonancelike) and interband (i.e., magnetoexcitonic) transitions are allowed both in single-layer graphene (SLG)\cite{AndoSLG04,Goerb07} and bilayer graphene (BLG)\cite{AndoBLG04}. Interband magnetophonon resonance (MPR) has indeed been observed in magneto-Raman scattering on SLG on the surface of graphite\cite{Yan10,Faug11} and non-Bernal stacked multilayer graphene on SiC\cite{Faug09}.

Graphite, a semimetal containing both electrons and holes even at zero temperature, is expected to exhibit even richer carrier-phonon coupling phenomena. Indeed, Ref.~\onlinecite{Kossa} recently reported magneto-Raman measurements on graphite up to 28~T, and observed inter-LL transitions and signatures of MPR. As described via the Slonczewski-Weiss-McClure (SWM) model\cite{Slo58,McC57,McC60}, graphite has a linear (``massless'') dispersion for the hole pocket around the $H$ point of the Brillouin Zone and a parabolic (``massive'') dispersion for the electron pocket around the $K$ point.  Angle-resolved photoemission measurements provided evidence of such massless and massive quasiparticles in graphite\cite{ZhouAPRES06}.  Near these high symmetry points, graphite's band structure can be approximated as a combination of SLG, describing the $H$-point massless holes, and BLG, describing the $K$-point massive electrons\cite{KosAndo08}.

Here, we report low-temperature magneto-Raman measurements of natural graphite in a magnetic field ($B$) up to 45~T, a range of fields much broader than any previous study, to the best of our knowledge.  We demonstrate a rich picture of MPR effects caused by coupling of the $E_{2g}$ phonon to both $H$-point (SLG-like) and $K$-point (BLG-like) interband excitations. We also observe a series of {\em electronic} Raman excitations (i.e., emission of electron-hole pairs instead of phonons), including transitions involving the lowest, electron-hole mixed, LLs.  We explain the entire, complex set of Raman-active interband excitations within a SWM approach. Furthermore, through quantitative analysis of the observed anticrossing behaviors, we determine the strengths of electron-phonon coupling (EPC) for both $H$-point holes and $K$-point electrons. Finally, in the highest magnetic-field range ($>$35~T), where all transition energies are far away from the $E_{2g}$ phonon energy, the $G$ peak narrows, due to suppression of the EPC contribution to the linewidth.

Raman spectra were collected on natural graphite (NGS Naturgraphit GmbH) in a backscattering geometry, with 
$B$ up to 45~T [see Fig.~1(a)].  A 532-nm laser is coupled via an optical fiber to the low-temperature probe, and focused to a spot of $\lesssim20~\mu$m, with a power of $\sim$13~mW. The probe is inserted into a helium cryostat and placed in a 31-T resistive magnet or 45-T hybrid magnet. Under  laser illumination, the temperature of the sample is stabilized at $\sim$10~K. The unpolarized Stokes component of the scattered light is directed into the collection fiber and guided to a spectrometer equipped with a charge-coupled-device camera. Most of the data were collected with a spectral resolution of $\sim$3.4~cm$^{-1}$. However, we used a spectral resolution $\sim$0.5~cm$^{-1}$ to accurately measure the full width at half maximum of the $G$ peak, FWHM(G), at selected magnetic fields between 32~T and 45~T. Raw data 
contains the signal of interest from the sample in a smooth background coming from the fibers.
 At frequencies $\gtrsim$1300~cm$^{-1}$, the background is featureless and much smaller than the signal from the sample. We performed numerous tests to characterize the field and temperature dependence of the background, in particular at frequencies $\lesssim$900~cm$^{-1}$, where parasitic scattering in fibers becomes comparable or higher than the signal from the sample. As the magnetic field increases up to 45T, we observe small broad-band changes reaching $\pm$1\% at $\sim$650 cm$^{-1}$. However, these could be due to other factors, such as long-term variation of the laser power or temperature drifts. Thus, we assume the background to be field independent, at least within our signal-to-noise ratio and use zero-field reference spectra to remove spurious signals due to scattering in the fibers. Decoupled SLG may exist on the surface of bulk graphite\cite{NeugOrlita09,LiAndrei10}. To avoid contributions from such SLGs, we recorded the spectra of several locations to select a region with a bulk graphitelike Raman spectrum, as indicated by the 2D Raman peak shape\cite{FerrarietAl06PRL,Ferrari07ssc}.  This approach is opposite to that of Refs.~\onlinecite{Yan10} and \onlinecite{Faug11}, where the samples were scanned to find SLG Raman signatures.

\begin{figure}
\includegraphics[scale=0.95]{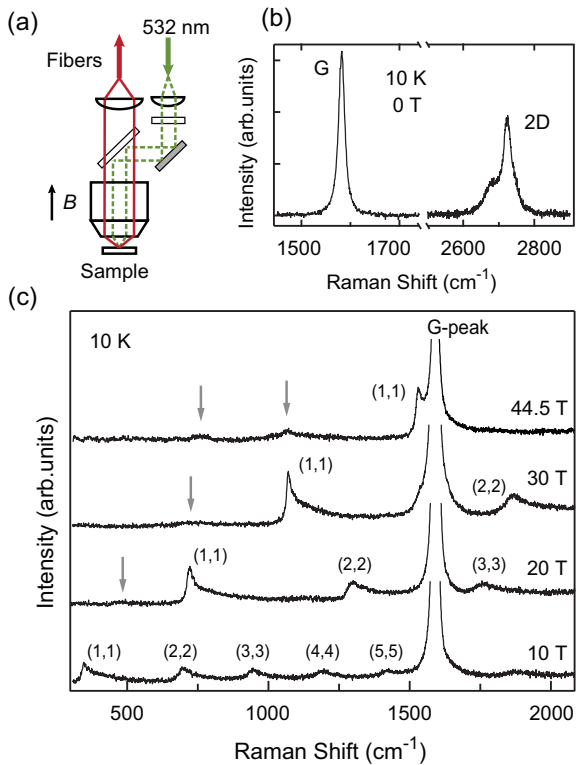}
\caption{(Color online) (a) Experimental configuration.  (b) Unpolarized, background-corrected Raman spectrum of natural graphite at zero magnetic field and 10~K.  (c) Magneto-Raman spectra at 10~K for various magnetic fields.  A series of interband electronic transitions are observed, as indicated, as well as a large $G$ peak$\sim$1580~cm$^{-1}$.  Gray arrows point to weak low-frequency scattering peaks.}
\end{figure}

At $B=0$, the first-order Raman spectrum of graphite is dominated by the $G$ peak at $\sim$1580~cm$^{-1}$, as shown in Fig.~1(b), due to scattering by the doubly degenerate zone-center $E_{2g}$ phonon\cite{TK70,Ferrari07ssc}. As $B$ increases, a number of peaks emerge, as shown in Fig.~1(c). These peaks become sharper and move towards higher frequencies with increasing $B$. Similar peaks were previously reported in Ref.~\onlinecite{GarciaFlores09}, where Raman scattering of bulk graphite was measured in magnetic fields up to 6.5~T, but assigned to LLs in BLG.

Figure 2 displays a set of spectra taken at 10~K as a function of $B$ up to 45~T.  The observed nearly linear $B$ dependence suggests these features to be related to inter-LL excitations of massive carriers in the vicinity of the {\it K} point.  The most intense peaks are attributed to the so-called ``symmetric'' inter-LL excitations, \textit{hn $\rightarrow$en} or ($n$,$n$), i.e., the transitions from the \textit{n}th hole to the \textit{n}th electron LLs\cite{Nak76}. Indeed, 
Ref.~\onlinecite{Kashuba2009} and \onlinecite{MarcinFalko10}
theoretically showed that symmetric inter-LL excitations are Raman active in both SLG and BLG. These symmetric transitions were previously observed and analyzed through an effective BLG model\cite{Faug11}.

\begin{figure}
\includegraphics[scale=0.9]{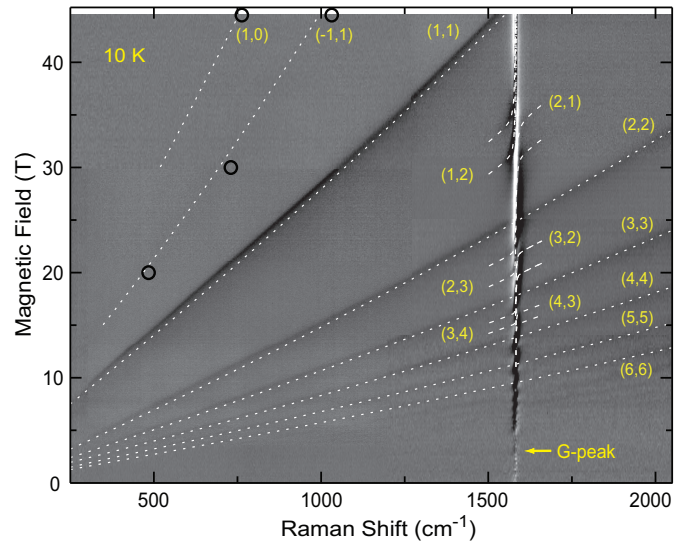}
\caption{(Color online) Intensity map of magnetic-field-dependent Raman intensity. Dotted lines represent the calculated energies of Raman-active, $hn$ $\rightarrow$ $en$, electronic excitations near the $K$ point.  Dashed lines represent asymmetric, $hn$ $\rightarrow$ $e(n-1)$  and $hn$ $\rightarrow$ $e(n+1)$ $K$ point  excitations coupled to the $E_{2g}$ phonon.  Open circles indicate the weak low-frequency peaks labeled by gray arrows in Fig.~1(c).}
\end{figure}

In addition, we detect two extra electronic features below the (1,1) transition, indicated by open circles in Fig.~2.  They are resolved at 45~T, as shown by gray arrows in Fig.~1, although their intensity is less than 10$\%$ of the (1,1) peak.  We attribute them to the lowest inter-LL transitions, (1,0) and ($-$1,1), at the {\it K} point.  They can be considered as a special case of the weak lowest-energy Raman-active transition in BLG predicted in Ref.~\onlinecite{MarcinFalko10}.

To validate our peak assignments, we calculate the energies of interband, inter-LL transitions within the SWM model.  This has seven tight-binding parameters, $\gamma _{0}$ to $\gamma _{5}$ and $\Delta$.  Despite its extensive use over the past 50 years, the precise values of these parameters are still under debate. Without the trigonal warping effect represented by $\gamma_{3}$, each LL can be obtained through a $4\times4$ Hamiltonian, which can be diagonalized for each $n$. Adding the $\gamma_{3}$ term mixes different LLs with indices $n$ and $n\pm3$, making the dimension of the Hamiltonian infinite. We numerically calculate the LL energies by truncating this Hamiltonian into a finite $\sim400 \times 400$ matrix. We note that, for matrix sizes larger than $100\times100$, the gaps between energy levels at $B=$10~T change less than 0.1~cm$^{-1}$. For higher magnetic fields, the results converges even faster. We obtained $\gamma_0$ from the position of the {\it H}-point MPR and used the SWM parameters from Ref.~\onlinecite{JarHer11} as the initial guesses for our fitting. To reduce the number of parameters, we fixed $\gamma_0$ and $\gamma_2$ and varied the others to fit the data.

Table I compares our results with values extracted from magnetotransport experiments\cite{JarHer11}, infrared magnetoreflectance spectroscopy\cite{Tung11}, magneto-Raman measurements\cite{Kossa}, as well as values deduced from earlier infrared magneto-spectroscopy experiments\cite{Brandt88}.  Though the tight-binding parameters are not significantly different, our spectroscopic observation of both symmetric and asymmetric transitions, including the low-energy transitions involving the electron-hole mixed $-$1 and 0 LLs, enables an accurate determination of the SWM parameters.

\begin{table}
\caption{\label{table} SWM band parameters (in eV) extracted from results in Fig.~2, in comparison with previously reported values.}
\begin{ruledtabular}
\begin{tabular}{lccccc}

&		This work	&	 Ref.~\onlinecite{JarHer11}	&	Ref.~\onlinecite{Tung11}	&		Ref.~\onlinecite{Kossa}	&	Ref.~\onlinecite{Brandt88}\\
\hline
$\gamma_{0}$			&	3.06 (1)		&	3.1		&	3.18 (3)	&	3.08 (1)		&	3.16 (5)\\
$\gamma_{1}$			&	0.370 (5)		&	0.39		&	0.38 (1)	&	0.380 (2)		&	0.39 (1)\\
$\gamma_{2}$			&	-0.028 (4)		&	-0.028 (4) 	&	-0.02 	&-\footnotemark[1]	&	-0.020 (2)\\
$\gamma_{3}$			&	0.33 (1)		&	0.315	&	-		&	0.315 (1)		&	0.315 (15)\\
$\gamma_{4}$			&	0.080 (5)		&	0.041 (10)	&	0.08 (3)	&	0.044 (5)		&	0.044 (24)\\
$\Delta+2\gamma_{5}$	&	0.130 (3)		&	0.15 (3)	&	0.064 (3)	&-\footnotemark[1]	&	0.084 (7)\\

\end{tabular}
\end{ruledtabular}
\footnotetext[1]{$\Delta+2\gamma_{5}-2\gamma_{2}$ = 0.22 (1)}
\end{table}

\begin{figure}
\includegraphics[scale=0.95]{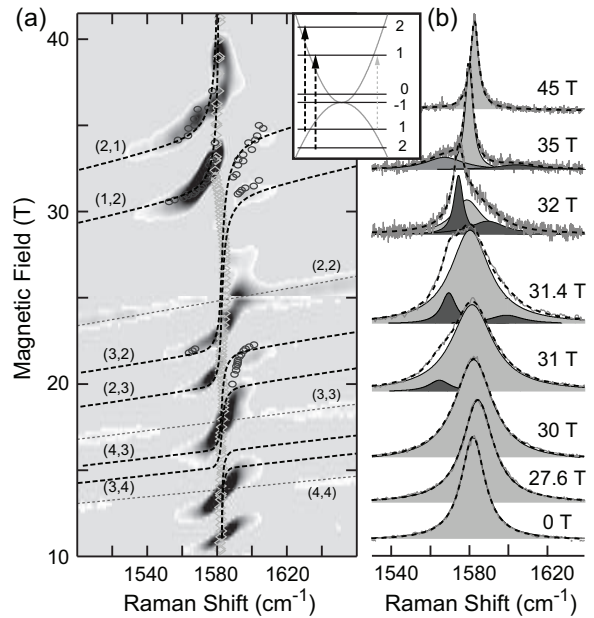}
\caption{Magnetophonon resonance at the {\it K} point. (a) Second derivative Raman intensity map.  Circles: Positions extracted from Lorentzian fits.  Dashed lines: Calculated energies of coupled {\it K}-electron-phonon modes. (b) $G$ peak measured at different $B$, with multiple-Lorentzian fits. Inset: schematic energy diagram and MPR-coupled inter-LL transitions at the {\it K} point.}
\end{figure}

Close examination of the $G$ peak in Fig.~2 reveals peak position modulations as a function of $B$.  At a certain $B$, the resonance condition $E_{n,n'}=\hbar\Omega_{\Gamma}$ is met, where $E_{n,n'}$ is the ($n$,$n'$) transition energy and $\Omega_{\Gamma}$ is the $E_{2g}$ phonon frequency, and the phonon is ``dressed'' by the electronic transition\cite{AndoSLG04,Goerb07,AndoBLG04}. This coupling manifests itself as a series of avoided crossings\cite{Faug09,Yan10,Faug11}.  Specifically, the $E_{2g}$ phonon is allowed to couple with an ($n$,$n'$) transition only when $|n|-|n'|=\pm{1}$.  To examine the data more closely, we fit the $G$ peak with Lorentzians and plot the extracted peak positions and the second derivative Raman intensity in Fig.~3(a).  The data reveals anticrossings at 34, 31, 21, and 19~T, corresponding to the (2,1), (1,2), (3,2), and (2,3) transitions, respectively. At lower fields, the doublet structure due to the (3,4) and (4,3) transitions is smeared out and appears as a weak modulation of the $G$ peak.  Note that, when the symmetric ($n$,$n$) peaks cross the $G$ peak, they appear unchanged, indicating the absence of coupling.  Furthermore, the central position of the $G$ peak is also $B$ dependent, exhibiting a modulation and broadening at $\sim$30~T (Fig.~4), which we interpret as a signature of MPR of the asymmetric  $h1$ $\rightarrow$ $0$ and $h1$ $\rightarrow$ $-1$ \textit{H}-point excitation with the $E_{2g}$ phonon. Finally, above 35~T,  where  the decay of $E_{2g}$ phonons into electron-hole pairs is quenched by Landau quantization and electron-phonon interaction is suppressed, the $G$ peak  narrows to $\sim$4.4~cm$^{-1}$. Our high-field value FWHM(G) is about twice the phonon-lifetime-limited linewidth at $B$ = 0, $\gamma_{\Gamma}^{ph-ph}\approx2.5~$cm$^{-1}$ (Refs.~\onlinecite{Bonini07,Chatzakis11}), indicating the presence of another, probably disorder-induced, broadening mechanism.

To analyze the observed MPR, we first focus on the $G$-peak sidebands, corresponding to coupled electron-phonon modes associated with {\it K}-point electron asymmetric transitions [Fig.~3(a)].  The doublet anticrossings at 34 and 31~T, corresponding to the (2,1) and (1,2) transitions, respectively, is most accurately resolved [Fig.~3(b)], and therefore, most suitable  for quantitative analysis.  Following Refs.~\onlinecite{Goerb07,Yan10},  we analyze the data via a two-coupled-mode model,
\begin{equation}
E_\pm=\frac{E_{G}+E_{n,n'}}{2}\pm\sqrt{\left(\frac{E_{G}-E_{n,n'}}{2}\right)^ 2+\mathnormal{g}^2},
\end{equation}
where $E_{G} = \hbar\Omega_{\Gamma} - i\gamma_{\Gamma}/2$, $E_{n,n'} = \hbar\Omega_{n,n'} - i\gamma_{n,n'}/2$, $\gamma_{\Gamma}$ ($\gamma_{n,n'}$) is FWHM(G) [($n$,$n'$) transition], and $g$ is the coupling parameter.  Expressing the magnetic energy $\hbar\omega_{B}$ at the {\it K} point within an effective BLG model, the coupling parameter $g$ is given by:
\begin{equation}
g^{(K)} =  \sqrt{\frac{\lambda_{\Gamma}^{(K)}}{4\pi}}\hbar\omega_{B} = \frac{3}{4} \sqrt{\frac{\lambda_{\Gamma}^{(K)}}{4\pi}} \frac{\gamma_{0}^{2}a^{2}}{\gamma_{1}l_{B}^{2}} \equiv g_0^{(K)}B,
\end{equation}
where $a$ (= 2.46~\AA) is the graphite lattice constant, $\gamma_0$ and $\gamma_1$ are tight binding parameters (see Table I), and $l_B=\sqrt{\hbar/eB}$ is the magnetic length. The dimensionless EPC $\lambda_{\Gamma}$ is defined following the notation of Refs.~\onlinecite{Basko09,Lazzeri06}:
\begin{equation}
\lambda_{\Gamma} = \frac{2 A_{u.c.}}{M\hbar\Omega_{\Gamma}}\frac{\langle{{D_{\Gamma}}^{2}}\rangle}{v_{F}^{2}}=
\frac{4}{\sqrt{3}} \frac{{\hbar}^2}{M\hbar\Omega_{\Gamma}} \frac{\langle{{D_{\Gamma}}^{2}}\rangle}{\gamma_0^{2}}
\end{equation}
where $A_{u.c.}$ is the graphene unit-cell area, $M$ is the carbon atomic mass, and $v_{F} = \frac{\sqrt{3}}{2\hbar}a\gamma_{0}$ = 0.99~$\times$~10$^{6}$~m/s is the Fermi velocity. $\langle{{D_{\Gamma}}^2}\rangle$ is the deformation potential of the $E_{2g}$ phonon, which describes the modulation of the coupling energy $\gamma_0$ by C-C bond length variation.

\begin{figure}
\includegraphics[scale=0.95]{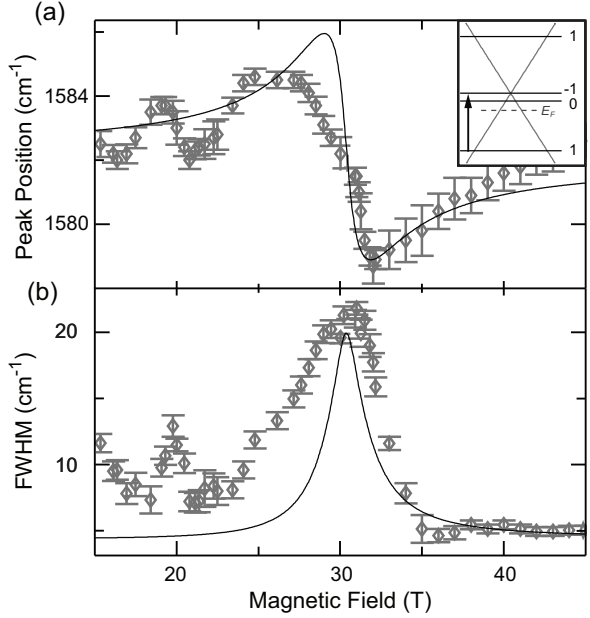}
\caption{Magnetophonon resonance at the $H$ point. (a) Peak position and (b) FWHM of the $G$-peak central Lorentzian component as a function of magnetic field. The model results of MPR-induced modulation are shown by solid lines. Inset: schematic energy diagram and MPR-coupled inter-LL transitions at the $H$ point.}
\end{figure}

The position and linewidth of unperturbed phonons can be derived from the high-field ($>$35~T) spectra, i.e. Pos(G) = 1582.6~cm$^{-1}$ and FWHM(G) = 4.4~cm$^{-1}$. We calculate the energies of asymmetric inter-LL transitions using the SWM parameters described previously. Fitting Eq.~(1) to the anticrossings at 31 and 34~T yields $\gamma_{\Gamma}=44\pm6$~cm$^{-1}$  and $g^{(K)}=0.72\pm0.03$~cm$^{-1}$/T.  We can thus extract $\lambda_{\Gamma}^{(K)}\approx 3.3\times10^{-2}$, in excellent agreement with that previously derived from density functional theory, the zero-field FWHM(G), the doping dependence of Pos(G), and the slope of the phonon dispersions around $\Gamma$: $\lambda_{\Gamma}^{(K)} \approx 3\times10^{-2}.$\cite{PiscanecetAl04PRL,PisanaetAlNatMat07,Yan07,Basko09}

Finally, we analyze the $B$-induced modulation of the central component of the $G$ peak, shown in Fig.~4.  The total peak-position modulation is $\sim$6~cm$^{-1}$, while the FWHM increases more than twice at $\sim$30~T.  This is consistent with MPR due to $H$-point inter-LL transitions, (1,0) or (1,$-$1), assuming that the LL widths are larger than the coupling strength. The $G$-peak modulation at $\sim$20~T is a signature of the MPR effect involving (2,3) $K$-point excitations. To deduce the EPC strength for the $H$-point, we model the 30~T resonance with Eq.~(1) using a SLG-like expression for $g^{(H)}$:

\begin{equation}
g^{(H)} = \sqrt{\frac{3}{2}}\sqrt{\frac{\lambda}{4\pi}}\frac{a}{l_{B}}\gamma_{0}\equiv g_0^{(H)}\sqrt{B}.
\end{equation}

The right-hand side of the resonance ($B>$~30~T) fits well with the model with $\gamma_{\Gamma}=100\pm10~$cm$^{-1}$, $g_0^{(H)}=3.2\pm0.2$~cm$^{-1}$/T$^{1/2}$, and $\lambda_{\Gamma}^{(H)} \approx 1.6$ $\times$ 10$^{-3}$. The discrepancy at lower fields is likely be due to the $E_{2g}$ renormalization via interaction with multiple inter-LL excitations, which cannot be spectrally resolved for $B$ $<$ 30~T.  We note that $\lambda_{\Gamma}^{(H)}$ is almost 20 times smaller than $\lambda_{\Gamma}^{(K)}$.

In summary, we performed high-field magneto-Raman experiments on graphite, observing strong magnetophonon resonances. The $G$ peak shifts and splits  as a function of magnetic field as it sequentially resonates with certain electronic transitions.  Analysis of the observed magnetophonon resonance effects allowed us to determine the strengths of electron-phonon coupling for both $H$- and $K$-point carriers. The Slonzcewski-Weiss-McClure model provides an accurate description of all observed interband electronic excitations. In the highest field range ($>$35~T), the $G$ peak narrows through reduced electron-phonon interaction.

This work was supported by NHMFL UCGP-5068 and DOE/BES DE-FG02-07ER46451.  J.K.~acknowledges support from DOE/BES through Grant No.~DE-FG02-06ER46308 and the Robert A.~Welch Foundation through Grant No.~C-1509.  Y.M.~and A.I.~acknowledge support from the Texas NHARP Grant No.~01889 and  A.~P.~Sloan Foundation.  A.C.F.~acknowledges funding from EPSRC Grants No. GR/S97613/01and No.EP/E500935/1, ERC grant NANOPOTS, EU-FP7 Grants GENIUS and RODIN, a Royal Society Wolfson Research Merit Award.  The measurements were carried out at the National High Magnetic Field Laboratory, which is supported by NSF Cooperative Agreement No.~DMR-0654118, by the State of Florida, and by the DOE.

\end{document}